\documentclass{elsart}

\usepackage{harvard,graphicx}

\usepackage{amssymb}

\def\url#1{{\ttfamily\def\/{/\discretionary{}{}{}}#1}}

\begin{document}

\begin{frontmatter}
\title{The Hierarchical Origin of Galaxy Morphologies}
\author{Matthias Steinmetz\thanksref{email1}}
\address{Steward Observatory, 933 N Cherry Ave, Tucson, AZ 85721, USA}

\author{Julio F. Navarro\thanksref{email2}}
\address{Department of Physics and Astronomy, University of Victoria, Victoria, BC, V8P~1A1, Canada}

\thanks[email1]{Alfred P.~Sloan Fellow \& Packard Fellow. E-mail: msteinmetz@as.arizona.edu}
\thanks[email2]{CIAR Scholar \& Alfred P.~Sloan Fellow. E-mail: jfn@uvic.ca}


\begin{abstract}
We report first results from a series of N-body/gasdynamical
simulations designed to study the origin of galaxy morphologies in a
cold dark matter-dominated universe. The simulations include star
formation and feedback and have numerical resolution sufficiently high
to allow for a direct investigation of the morphology of simulated
galaxies. We find, in agreement with previous theoretical work, that
the presence of the main morphological components of galaxies--disks,
spheroids, bars--is regulated by the mode of gas accretion and
intimately linked to discrete accretion events. In the case we
present, disks arise from the smooth deposition of cooled gas at the
center of dark halos, spheroids result from the stirring of
preexisting disks during mergers, and bars are triggered by tides
generated by satellites. This demonstrates that morphology is a
transient phenomenon within the lifetime of a galaxy and that the
Hubble sequence reflects the varied accretion histories of galaxies in
hierarchical formation scenarios.  In particular, we demonstrate
directly that disk/bulge systems can be built and rebuilt by the
smooth accretion of gas onto the remnant of a major merger and that
the present-day remnants of late dissipative mergers between disks are
spheroidal stellar systems with structure resembling that of field
ellipticals. The perplexing variety of galaxy morphologies is thus
highly suggestive of--and may actually even demand--a universe where
structures have evolved hierarchically.
\end{abstract}

\end{frontmatter}

\section{Introduction}

The bewildering variety of galaxy morphologies has long marveled and
challenged astronomers (Hubble 1926). Disks, bulges, halos, bars,
tails; accounting for these basic features of galaxy taxonomy has
elicited much controversy in the past, but is now widely ascribed to
the particular history of mergers and accretion events that galaxies
experience during their assembly in a hierarchical universe (Kauffmann
et al.\ 1993, Baugh et al.\ 1998). In such scenarios, one example of
which is the popular cold dark matter (CDM) cosmogony, disks are
envisioned to form as the result of gas accreted smoothly from the
intergalactic medium (see, e.g., Katz \& Gunn 1991, Navarro \& White
1994, Steinmetz \& M\"uller 1994), 
whereas spheroids are the remnants of major merger events where
disks are thrown together and mixed violently on a short timescale
(Toomre 1977, Barnes \& Hernquist 1992). Galaxy morphology thus
evolves continuously throughout a galaxy's lifetime, and is determined
by a delicate balance between the mode of gas accretion and the
detailed merger history of an individual galaxy. Here we present a
cosmological gasdynamical simulation which confirms that a single
galaxy in the CDM cosmogony may run through the whole Hubble sequence
during its lifetime, validating previous theoretical expectation and
illustrating the inextricable link between morphology and the
hierarchical mode of galaxy formation.

\section{The Simulation}

The continual metamorphosis of galaxy morphologies in hierarchically
clustering universes is well illustrated by the evolutionary sequence
shown in Figures 1 to 5. These figures show, at various times, the
`luminous' (baryonic) component of a massive dark halo
($\sim2.5\times10^{12}$ M$_\odot$ at z=0) formed in the $\Lambda$CDM
cosmogony (Bahcall et al.  1999). $\Lambda$CDM assumes a low-density
universe ($\Omega_0 = 0.3$), currently dominated by a cosmological
constant ($\Omega_\Lambda = 0.7$), with a baryonic density parameter
$\Omega_b=0.019\,h^{-2}$, and a Hubble parameter $h =
H_0/(100\,\mbox{km\,s$^{-1}$\,Mpc$^{-1}$})=0.65$. The $\Lambda$CDM
power spectrum is normalized so that the present-day rms mass
fluctuations on spheres of radius 8 $h^{-1}$ Mpc is $\sigma_8=0.9$.

The evolution of this system is characterized by episodes of smooth
accretion punctuated by two major merger events, one at z$\sim$3.3 and
the second at z$\sim$0.6. The simulations are performed with GRAPESPH
(Steinmetz 1996), a cosmological hydrodynamics code that includes the
effects of gravity, gas dynamics, and radiative cooling and heating
processes. Star formation is included via a phenomenological recipe
that converts unstable gas regions into stars at rates chosen to match
observational constraints (Steinmetz \& Navarro 1999). The simulation
concentrates computational resources on the surroundings of a single
dark halo identified at z=0, but it includes nonetheless the full
tidal field of a large, representative volume of the $\Lambda$CDM
cosmogony (see, e.g., Navarro \& White 1994).  The particle mass in
the simulation is $1.27\times 10^7\,$M$_{\odot}$ for the gas component
and $5.76\times 10^{7}\,$M$_\odot$ for the dark matter component,
respectively. The Plummer gravitational softening is fixed at $0.5$
(physical) kpc.

\begin{figure}
\begin{center}
\includegraphics*[width=13.5cm]{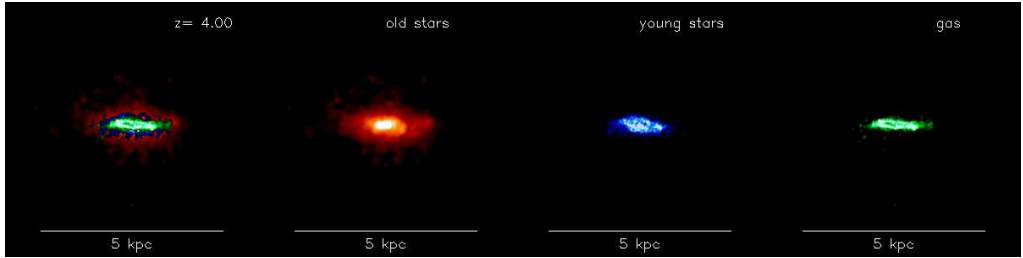}
\end{center}
\caption{The
most massive progenitor at z=4 shown edge-on. Gas particles are shown in
green, `young' (i.e. less than 200 Myr old) stars in blue and older stars
in red. Horizontal bars in each panel are 5 (physical) kpc long.}
\end{figure}

\section{Disks and Bulges} 

As early as z=4, the most massive progenitor of the galaxy has already
assembled $\sim3\times10^{10}$ M$_{\odot}$ of gas and stars in a
disk-like configuration, as shown in Figure 1. The disk is rather
small, about 3 (physical) kpc in diameter, and rotates at
approximately 180 km s$^{-1}$. In agreement with the main premise of
the hierarchical scenario, the disk is the outcome of smooth accretion
of gas and of its progressive transformation into stars over the
lifetime of the system.  Star formation begins in earnest at z$\sim$10
and proceeds at an average rate of $\sim$30 M$_{\odot}$ yr$^{-1}$,
albeit spread over a number of subsystems arranged along a filamentary
structure. The largest of the progenitor subsystems is illustrated in
Figure 1. Stars younger than 200 Myr are shown in blue, older stars in
red.  The ``pure disk" phase in the evolution of this galaxy comes to
an end rather abruptly at z$\sim$3.3, when the galaxy merges with a
similarly-sized companion. The merger stirs the stars from the disk
components and assembles them into a single spheroidal component---the
progenitor of a bulge (see the z=3.15 panel in Figure 2). The merger
is also accompanied by a burst of star formation that largely depletes
the gas content of the merging disks, transforming $1.6\times10^{10}$
M$_{\odot}$ of gas into stars in just 300 Myr. 

Just before the burst, the apparent brightness of the galaxy is about
$26.5$ mag in the $r$-band (assuming $1.5$ mag of dust dimming), about
$1$ mag fainter than the current limiting magnitude of the
spectroscopically-confirmed `Lyman-break' population (Steidel et al.\
1996). During the burst at $z=3.3$, however, its maximum brightness
reaches $25$ $r$-mag, compatible with that in current Lyman-break
spectroscopic samples. This implies that at least some of the bright
Lyman-break systems may be the result of merger-triggered starbursts
associated with the epoch of bulge formation (Somerville et al.\
2000). We note, however, that overall the brightening during the burst
is modest, and that this galaxy is not a dwarf system: its stellar
mass at $z=3.3$ is $6.3\times10^{10}\, M_{\odot}$, comfortably in the range of best
estimates available for the Lyman-break galaxies (Papovich et al. 2001, Rudnick
2001, Shapley et al. 2001).  This result,
together with the short duty cycle associated with this event (about
300 Myr) and the strong clustering of bright Ly-break systems,
supports early theoretical suggestions that a substantial fraction of
the Lyman-break galaxies may inhabit massive halos and may be the
progenitors of bright cluster ellipticals (Baugh et al.\ 1998, Mo, Mao
\& White 1999). 

\begin{figure}
\begin{center}
\includegraphics*[width=13.5cm]{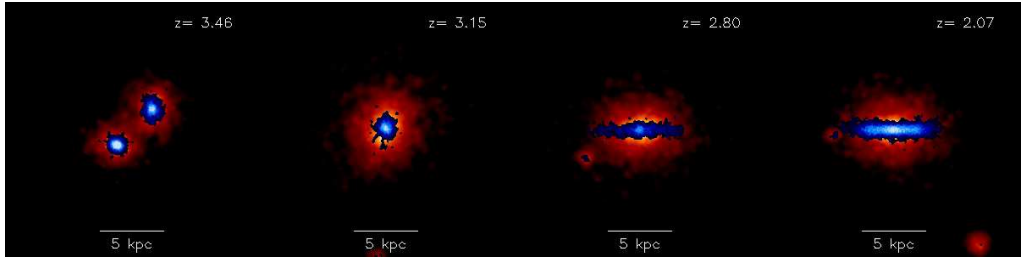}
\end{center}
\caption{The formation of a bulge and the rebirth of a disk.  The
sequence shows the formation of a bulge by the merger of two almost
equal mass `pure disk' systems at z$\sim$3. After z=3 smooth accretion
of gas regenerates the disk component around the bulge. Blue is used
in each panel to denote stars formed since the preceding frame (since
z=4.2 for the first panel), red for older stars. Horizontal bars in
each panel are 5 (physical) kpc long.}
\end{figure}

\section{The Rebirth of a Disk Galaxy.}

From $z=3$ to $z=1.8$ (see Figure 2) the baryonic mass of the galaxy
increases by {$\sim50$\%} through smooth accretion of intergalactic
gas not attached to massive clumps. As expected, the accreting gas
radiates its energy quite efficiently and assembles into a new,
distinct disk component where star formation proceeds at a rate that
declines from 20 M$_{\odot}$ yr$^{-1}$ at z$\sim$3 to 8 M$_{\odot}$
yr$^{-1}$ at z=1.8. The disk is clearly assembled from the inside out,
although there is substantial scatter in the angular momentum of the
accreting gas and non negligible amounts of late-accreting, low
angular momentum gas continues to fuel star formation near the center
and throughout the disk. At z=1.8 the simulated galaxy resembles a
bright ($M=-22$ in $r$) Sa/Sb spiral (Figure 3). Over 70\% of the
rest-frame optical luminosity comes from a roughly exponential disk
with $\sim$1.5 (physical) kpc scale length while the rest comes from a
spheroidal component whose spatial distribution may be well
approximated by an r$^{1/4}$-law with a $\sim$1 kpc effective
radius. {\it This is to date the most direct demonstration that
rotationally supported disks of gas and young stars, as well as
dynamically hot, centrally concentrated bulge-halo systems of old
stars---the main luminous components of present-day disk galaxies---are
natural byproducts of the hierarchical assembly of a galaxy.}

\begin{figure}
\begin{center}
\includegraphics*[width=13.5cm]{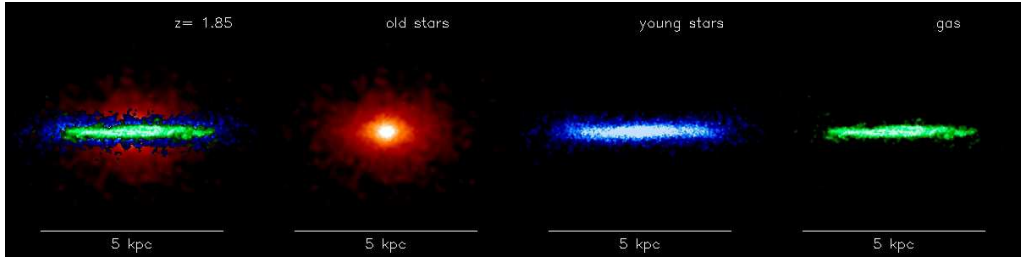}
\end{center}
\caption{The appearance of the
galaxy at z=1.8, seen edge-on. Green is used for the gas, blue for `young'
stars (i.e. formed since z=3), red for older stars. At this time, the
morphology of the galaxy is reminiscent of early type spirals, with a
dense bulge surrounded by a disk of gas and young stars. Horizontal bars in each
panel are 5 (physical) kpc long.}
\end{figure}

\section{From Disk Galaxy to Barred Galaxy} 

Between z=1.8 and z=0.7 the disk undergoes a further dramatic
transformation as the dominant mode of gas accretion switches from smooth
buildup of intergalactic gas to the accretion of discrete satellites. The
most significant of these events adds $\sim$10\% of mass to the spheroidal
component at z=1.18, when a close passage through the disk plane violently
disrupts the satellite (few of the satellite's stars make it into the
disk), in a process highly reminiscent of the ongoing disruption of the
Sagittarius dwarf by the Galaxy (Ibata et al.\ 1994, Helmi et al.\ 1999).
The satellite's main effect on the disk morphology, however, dates to its
first pericentric (10 kpc) passage, at z=1.62, when the satellite's tide
forces a distinct bar pattern on the disk, clearly seen in 
Figure 4. 

The bar pattern persists long after the disruption of the culprit
satellite and its response is strongest in the gas, leading to the
formation of young stars (shown in blue) that trace the bar pattern
better than older disk stars (in red). The bar extends out to about
2.5 kpc and has a corotation radius of slightly less than 3 kpc,
implying, in agreement with the few barred galaxies where this ratio
has been measured (Debattista \& Sellwood 1999), that the bar is
``fast".  The baryonic component dominates the mass distribution near
the center (75\% of the mass within 3 kpc is in gas and stars) and the
bar pattern appears stable, clearly lasting for more than 30 orbital
periods. Although the process described here is likely not the only
bar formation mechanism acting on real galaxies, our simulation shows
nevertheless a direct link between the presence of a bar pattern and
the existence of satellites that may act as tidal triggers. It thus
suggests that substantial insight into the accretion history of the
galaxy population may be gained from studying the frequency and
strength of barred patterns in spiral galaxies as a function of
redshift, as highlighted in recent work (Abraham et al.\ 1999).

\begin{figure}
\begin{center}
\includegraphics*[width=13.5cm]{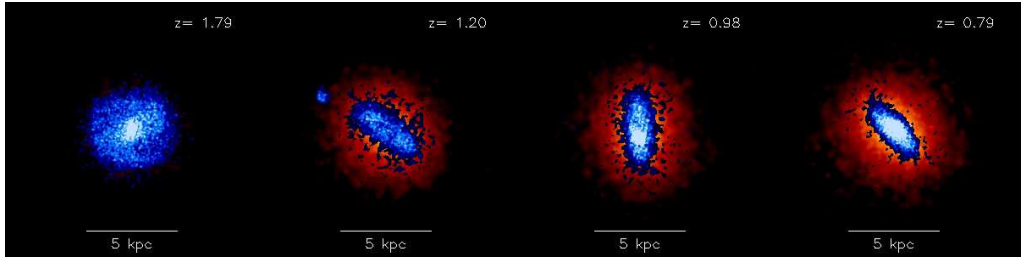}
\end{center}
\caption{The tidal triggering of bar instability by a satellite. At
z=1.6 the disk develops a well defined, long-lasting bar pattern as a
result of tidal forcing by the satellite shown in the z=1.2 panel. The
satellite is finally disrupted at z=1.18 after 7 pericentric
passages. The bar pattern, however, survives for several Gyr, as shown
in the picture. Stars less than 1.5 Gyr old are shown in blue, older
stars in red. Note that the bar is most prominent in the young
component. Horizontal bars in each panel are 5 (physical) kpc long.}
\end{figure}

\section{Ellipticals as the Outcome of Major Mergers}

The bar pattern in the disk is still quite strong at z=0.7, when the
galaxy undergoes a major merger with a companion roughly half as
massive.  The outcome of such major merger---a spheroidal pile of
stars---has been discussed extensively in the literature since the
Toomres first suggested mergers of spirals as a possible route to
forming elliptical galaxies (Toomre \& Toomre 1972, Toomre 1977). As
illustrated in Figure 5, the final collision between the two main
bodies of the colliding galaxies occurs at z$\sim$0.6, and leaves
behind a triaxial stellar system reminiscent of nearby field
ellipticals (de Zeeuw \& Franx 1991). The remaining gas within the two
spirals ($\sim$10\% of the total baryonic mass) is efficiently
funneled to the center during the merger, where it is quickly
converted into stars (Hernquist \& Barnes 1991). This last episode of
star formation is effectively over by z=0.5, leaving at z=0.27
(rightmost panel of Figure 5) only a small ($\sim$1 kpc diameter)
`core' of metal-rich stars younger than 1.5 Gyr (in blue in Figure 5),
surrounded by a large spheroid of older stars (shown in
red). Exhausted its star formation fuel, stars in the galaxy evolve
passively for the remaining 3.5 Gyr. 

At z=0 the galaxy closely resembles an elliptical both dynamically and
in its stellar population. With a central (i.e. within 3 kpc) 1-d
velocity dispersion of 310 km s$^{-1}$ and a $\sim 1.3$ kpc effective
radius this $M_r=-22$ galaxy sits close to the `Fundamental Plane'
drawn by ellipticals at $z=0$ (Jorgensen et al.\ 1996), although its
radius seems significantly smaller than typical ellipticals of this
luminosity. The stellar body of the galaxy is mildly triaxial, and
rotates around its minor axis at a maximum speed of $\sim 200$ km
s$^{-1}$, with little indication of large misalignments between its
dynamical and structural axes. Its mean color, $B-V=1.0$ (Vega
magnitudes), is also consistent with those of bright spheroids in the
local universe, and even the presence of a slightly younger `core'
bears a striking resemblance to the detailed structure of field
ellipticals recently unveiled by Hubble Deep Field studies (Menanteau
et al.\ 2001).

\begin{figure}
\begin{center}
\includegraphics*[width=13.5cm]{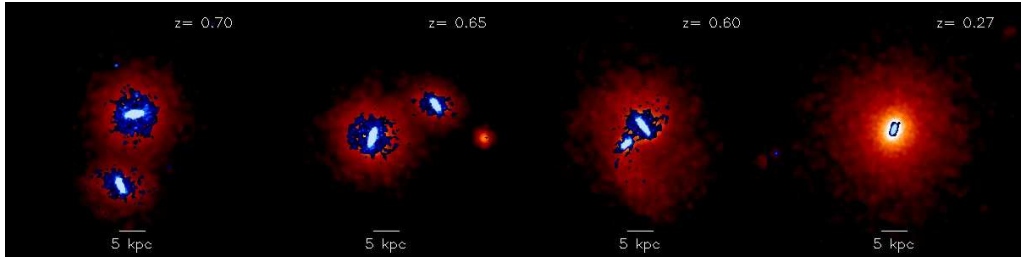}
\end{center}
\caption{A major merger and the formation of an elliptical galaxy. The
disk merges with another system of about half its mass at z=0.7 to
form a spheroidal system of stars resembling an elliptical galaxy. All
remaining gas is driven to the center of the remnant and consumed in a
burst of star formation lasting 600 Myr.  Stars less than 1.5 Gyr old
are shown in blue, older in red. A movie illustrating the whole
evolution of the galaxy can be found in the supplemental material
associated with this manuscript.}
\end{figure}

\section{Summary}

The evolutionary sequence portrayed in Figures 1 to 5 demonstrates
conclusively the ever-changing nature of galaxy morphology and the
intimate connection between morphology and accretion history. The
perplexing variety of observed galaxy morphologies thus arises
naturally in a hierarchical universe, where structures are assembled
through a combination of mergers and smooth accretion. Our study
confirms early theoretical work regarding the origin of all major
components making up present-day galaxies. In particular, the
simulation presented here shows that (i) most of the stars that make
up the halo and bulge of disk galaxies may have been born in early
proto-disks that were later stirred into spheroids by mergers, (ii)
that disk/bulge systems may be formed by smooth accretion of gas onto
the remnants of major mergers, (iii) that the population of
Lyman-break galaxies may include some ongoing merger events associated
with the formation of the first spheroids, and (iv) that the remnants
of late dissipative disk mergers are spheroidal stellar systems with
structure resembling that of field ellipticals.

These results provide clear support for the long-held view that the
Hubble sequence can be understood in terms of accretion histories and
that morphology is a transient phenomenon in the lifetime of a galaxy.
Our study also highlights some of the potential shortcomings of
hierarchical galaxy formation scenarios. For example, is the large
fraction of stars observed in disks today consistent with the `lumpy'
accretion histories expected in CDM-like scenarios? Can one account
for the observed frequency, size, and dynamical properties of `pure
disk' (bulge-less) galaxies ?  Can variations in accretion history
with environment account for the morphology-density relation?  These
questions are critical to the success of the hierarchical model and
remain challenging puzzles to be elucidated within this so far highly
successful paradigm. It will take a substantive computational effort
to address them through direct simulation, but one that is within
reach of today's technological capabilities. We look forward to a new
era when galaxy morphology will reach beyond its wondrous aesthetic
appeal and will fulfill its promise to decipher the intricate paths of
galaxy formation.

\section{Acknowledgments}

This work has been supported by the National Aeronautics and Space
Administration under NASA grants NAG 5-7151 and NAG 5-10827, 
NSF grant 9870151 and by NSERC
Research Grant 203263-98. MS and JFN are supported in part by fellowships
from the Alfred P.~Sloan Foundation. MS is also supported by a fellowship
from the David and Lucile Packard Foundation.

\end{document}